\documentclass[10pt,aps,prc,twocolumn,floatfix]{revtex4-1} 
\usepackage{graphicx}
\usepackage{sidecap}
\usepackage{array}
\usepackage{multirow}
\usepackage{subfigure}
\usepackage{color}
\usepackage{lineno}
\usepackage{float}               
\usepackage{placeins}
\usepackage{xspace}        
\usepackage{epsfig}
\usepackage{graphicx}
\usepackage{fancyhdr}   
\usepackage{pslatex}   
\usepackage{xcolor}
\usepackage{color}
\usepackage{amsmath, amsthm, amssymb} 
\usepackage[symbol]{footmisc}
\newcommand{\rootsnn} {\mbox{$\sqrt{s_{\rm NN}}$}\xspace}

\newcommand{\pt}{$p_{\mathrm{T}}~$}
\newcommand{\pta}{$p_{\mathrm{T}}$}

\begin{document}

\title{Non-monotonic energy dependence of net-proton number fluctuations}

\author{
J.~Adam$^{6}$,
L.~Adamczyk$^{2}$,
J.~R.~Adams$^{39}$,
J.~K.~Adkins$^{30}$,
G.~Agakishiev$^{28}$,
M.~M.~Aggarwal$^{41}$,
Z.~Ahammed$^{61}$,
I.~Alekseev$^{3,35}$,
D.~M.~Anderson$^{55}$,
A.~Aparin$^{28}$
% ...
}

% ...
\address{$^{2}$AGH University of Science and Technology, FPACS, Cracow 30-059, Poland}
\address{$^{3}$Alikhanov Institute for Theoretical and Experimental Physics NRC "Kurchatov Institute", Moscow 117218, Russia}
% ...
\address{$^{6}$Brookhaven National Laboratory, Upton, New York 11973}
% ...
\address{$^{28}$Joint Institute for Nuclear Research, Dubna 141 980, Russia}
% ...
\address{$^{30}$University of Kentucky, Lexington, Kentucky 40506-0055}
% ...
\address{$^{35}$National Research Nuclear University MEPhI, Moscow 115409, Russia}
% ...
\address{$^{39}$Ohio State University, Columbus, Ohio 43210}
% ...
\address{$^{41}$Panjab University, Chandigarh 160014, India}
% ...
\address{$^{55}$Texas A\&M University, College Station, Texas 77843}
% ...
\address{$^{61}$Variable Energy Cyclotron Centre, Kolkata 700064, India}
% ...

\collaboration{STAR Collaboration}\noaffiliation



\begin{abstract}
Non-monotonic variation with collision energy (\rootsnn) of the moments of the net-baryon number distribution in heavy-ion 
collisions, related to the correlation length and the susceptibilities of the system, is 
suggested as a signature for the Quantum Chromodynamics (QCD) critical point. We report the first evidence of a non-monotonic variation in
kurtosis times variance of the net-proton number (proxy for net-baryon
number) distribution as a function of  \rootsnn
with 3.1$\sigma$ significance, for head-on (central) gold-on-gold
(Au+Au) collisions measured  using the STAR
detector
at RHIC.  Data in non-central Au+Au
collisions and models of heavy-ion collisions without a critical point show a monotonic
variation as a function of \rootsnn. 
\end{abstract}
\maketitle
%

One of the fundamental goals in physics is to understand the
properties of matter when subjected to variations in temperature and
pressure. Currently, the study of the phases of strongly
interacting nuclear matter is the focus of many research activities
worldwide, both theoretically and experimentally~\cite{Bzdak:2019pkr,Luo:2017faz}.
The theory that governs the strong interactions is Quantum
Chromodynamics (QCD), and the corresponding phase diagram is
called the QCD phase diagram. From different examples of condensed-matter systems,
experimental progress in mapping out phase diagrams is achieved by
changing the material doping, adding more holes than electrons.
Similarly it is suggested for the QCD phase diagram, that adding more
quarks than antiquarks (the energy required is defined by
the baryonic chemical potential, $\mu_{\mathrm B}$), through changing
the heavy-ion collision energy, enables a search for new emergent
properties and a possible
critical point in the phase diagram.
The phase diagram of QCD has at least two distinct phases: a Quark
Gluon Plasma (QGP) at higher
temperatures, and a state of confined quarks and gluons at lower temperatures called the hadronic
phase~\cite{Fukushima:2010bq, BraunMunzinger:2009zz, Asakawa:1989bq}.
It is inferred from lattice QCD calculations~\cite{Aoki:2006we} that
the transition is consistent with being a cross over at small
$\mu_{\mathrm B}$, and that the transition temperature is about 155 MeV~\cite{Aoki:2009sc,Bazavov:2011nk,Gupta:2011wh}. 
An important predicted feature of the QCD phase structure is a
critical point 
~\cite{Stephanov:2008qz,Stephanov:1999zu}, followed
at higher $\mu_{\mathrm B}$ by a first order phase transition.
Attempts are being made to locate the predicted critical point both 
experimentally and theoretically. Current theoretical 
calculations are highly uncertain about the location of the critical 
point.  Lattice QCD calculations at finite  $\mu_{\mathrm B}$ face 
numerical challenges in computing~\cite{Bazavov:2017tot, Bazavov:2017dus}. Within these limitations, the 
current best estimate from lattice QCD is that if there is a critical
point, its location is likely above $\mu_{\mathrm B}$ $\sim$ 300
MeV~\cite{Bazavov:2017tot, Bazavov:2017dus}.
The goal of this work is to search for possible signatures of the
critical point by varying the collision energy in heavy ion collisions
to cover a wide range in effective temperature  ($T$) and
$\mu_{\mathrm B}$ in the QCD phase diagram~\cite{Stephanov:2011pb}.

Another key aspect of investigating the QCD phase diagram is to determine whether the system has attained thermal
equilibrium. Several theoretical interpretations of experimental data have the underlying assumption that
the system produced in the collisions should have come to local thermal equilibrium during
its evolution. Experimental tests of thermalization for
these femto-scale expanding systems are non-trivial.  However, the
yields of produced hadrons and fluctuations of multiplicity
distributions related to conserved quantities have been studied and
shown to have characteristics of thermodynamic equilibrium for higher
collision energies \cite{Bazavov:2017tot, Adamczyk:2017iwn,
  Andronic:2017pug, Almasi:2017bhq, Bazavov:2012vg, Borsanyi:2014ewa, Gupta:2020pjd}.

Upon approaching a critical point,
the correlation length diverges and thus 
renders, to a large extent, microscopic details irrelevant.  Hence 
observables like  the moments of the conserved net-baryon number
distribution, which are sensitive to the correlation length,
are of interest when searching for a critical point.
A non-monotonic variation of these moments as a function of \rootsnn has been proposed as 
an experimental signature of a critical point~\cite{Stephanov:1999zu,Stephanov:2011pb}. 
However, considering the complexity of
the system formed in heavy-ion collisions, signatures of a critical point are
detectable only if they can
survive the evolution of the system, including the effects of finite size and time~\cite{Berdnikov:1999ph}.
Hence, it was proposed to study higher moments of distributions of 
conserved quantities ($N$) 
due to their stronger dependence on the correlation length~\cite{Stephanov:2008qz}. 
The promising higher moments are the skewness, 
${\it {S}}$ = $\left\langle (\delta N)^3 \right\rangle/\sigma^{3}$,
and kurtosis, $\kappa$ = [$\left\langle (\delta N)^4 \right\rangle/\sigma^{4}$] -- 3,
where $\delta N$ = $N$ -- $M$, $M$
is the mean and $\sigma$ is the standard deviation.
The magnitude and the sign of the moments, which quantify the shape of the multiplicity
distributions, are important for understanding the critical point~\cite{Stephanov:2011pb,Asakawa:2009aj}. An additional
crucial experimental challenge is to measure, on an event-by-event basis,
all of the baryons produced within the acceptance of a detector~\cite{Kitazawa:2012at,Bzdak:2012ab,Bzdak:2012an}. However,
theoretical calculations 
have shown that the proton-number fluctuations can also
reflect the baryon-number fluctuations at the critical
point~\cite{Hatta:2003wn, Kitazawa:2012at}.

The measurements reported here are from Au+Au collisions
recorded by the STAR detector~\cite{Ackermann:2002ad} at RHIC from
the years 2010 to 2017. 
The data is presented for \rootsnn = 7.7, 11.5, 14.5, 19.6, 27, 39, 54.4,
62.4 and 200 GeV as part of phase-I of the Beam Energy Scan (BES)
program at RHIC~\cite{Adamczyk:2017iwn}. These \rootsnn values correspond to
$\mu_{\mathrm B}$ values ranging from 420 MeV to 20 MeV  at chemical
freeze-out~\cite{Adamczyk:2017iwn}. All
valid Au+Au collisions occurring within 60 cm (80 cm for \rootsnn =
7.7 GeV) of
the nominal interaction point along the beam axis are selected.
For the results presented here, the number of minimum bias Au+Au collisions
ranges between 3 million for \rootsnn = 7.7 GeV and
585 million at \rootsnn = 54.4 GeV. 
These statistics are
found to be adequate to make the measurements of the moments of the net-proton
distributions up to the fourth order~\cite{Pandav:2018bdx}. 
The collisions are further divided into centrality classes characterised by their impact 
parameter, which is the closest distance between the centroid of two
nuclei passing by. In practice, the impact parameter is determined
indirectly from the measured multiplicity
of charged particles other than protons ($p$) and
anti-protons ($\bar{p}$)
in the pseudo-rapidity range $|\eta|$ $<$ 1,
where $\eta=- \ln [ \tan (\theta/2)]$, with $\theta$ being the angle
between the momentum of the particle and the positive direction of the
beam axis. 
We exclude $p$ and $\bar{p}$ while classifying events based on
impact parameter specifically to avoid self-correlation
effects~\cite{Chatterjee:2019fey}.
The effect of self-correlation potentially arising due to the decay of
heavier hadrons into $p$($\bar{p}$) and other charged particles has
been checked to be negligible from a study
using standard heavy-ion
collision event generators, HIJING~\cite{Wang:1991hta} and
UrQMD~\cite{Bleicher:1999xi}.
The effect of resonance decays and 
the pseudo-rapidity range for centrality determination have been understood 
and optimized using model
calculations~\cite{Luo:2013bmi,Garg:2013ata}. 
The results presented
here correspond to two event classes: central
collisions (impact parameters $\sim$ 0-3 fm, obtained from the top 5\%
of the above-mentioned multiplicity distribution) and peripheral
collisions (impact parameters $\sim$ 12-13 fm, obtained from the
70-80\% region of the multiplicity distribution).

The protons and anti-protons are identified, along with their momenta, 
by reconstructing their tracks in the Time Projection Chamber (TPC) placed 
within a solenoidal magnetic field of 0.5 Tesla, and by measuring their 
ionization energy loss ($dE/dx$) in the sensitive gas-filled volume of
the 
chamber. The selected kinematic region for protons covers all
azimuthal angles for the rapidity range $|y| < 0.5$, where rapidity
$y$ is
the  inverse hyperbolic tangent
of the component of speed 
parallel to the beam direction in units of the speed of light. The precise measurement of $dE/dx$ with a resolution of 7\% in 
Au+Au collisions allows for a clear identification of protons up to 
800 MeV/$c$ in transverse momentum (\pta). The identification 
for larger \pt (up to 2 GeV/$c$, with purity above 97\%)
is made by a Time Of Flight detector (TOF)~\cite{sup} having a
timing resolution of better than 100
ps.  A minimum \pt threshold of 400 MeV/$c$ and a maximum distance of 
closest approach to the collision vertex of 1 cm for each
$p$($\bar{p}$) candidate track is used to suppress contamination
from secondaries and other backgrounds~\cite{Adamczyk:2017iwn,Adamczyk:2013dal}.  
This \pt acceptance accounts for approximately 80\% of the total $p$ + $\bar{p}$
multiplicity at mid-rapidity. This is a significant improvement from
the results previously 
reported~\cite{Adamczyk:2013dal} which only had the  $p$ +
$\bar{p}$ measured using the TPC.
The
observation of non-monotonic variation of the kurtosis times
variance ($\kappa\sigma^{2}$) with energy is much more significant
with the increased acceptance. For the rapidity
dependence of the observable see Supplemental Material~\cite{sup}.

\begin{figure}[t!]
{\centerline{\includegraphics[scale=0.3]{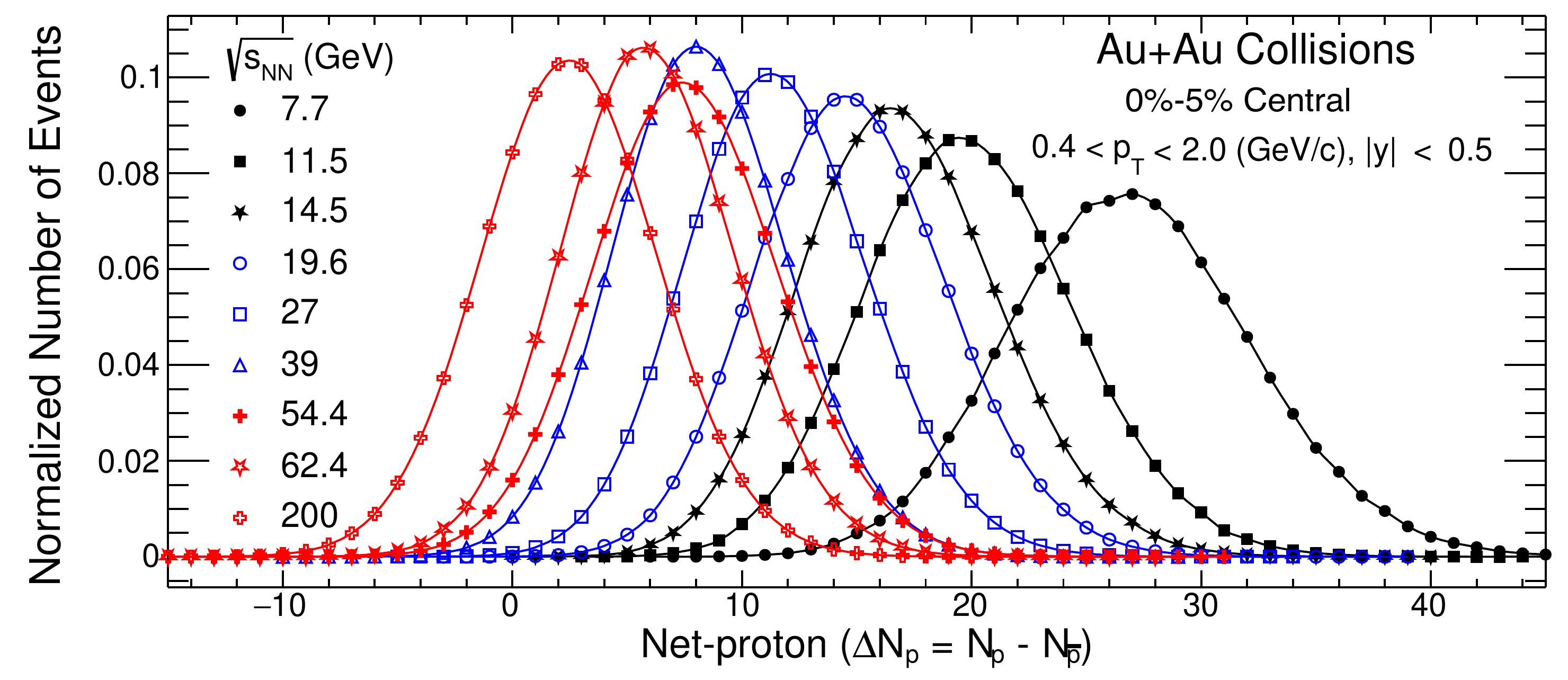}}}
\caption{
\label{fig:ebyedist}
Event-by-event net-proton number distributions for head-on (0-5\%
central) Au+Au collisions for nine \rootsnn values measured by
STAR. The distributions are normalized to the total
number of events at each \rootsnn.
The statistical uncertainties are smaller than the symbol sizes and
the lines are shown to guide the eye. The
distributions in this figure are not corrected for proton and
anti-proton detection efficiency. The deviation of the
  distribution for \rootsnn  = 54.4 GeV from the general energy
  dependence trend is understood to be due to the reconstruction
  efficiency of protons and anti-protons being different compared to
  other energies.
}
\end{figure}

Figure~\ref{fig:ebyedist} shows the event-by-event net-proton ($N_{p} - N_{\bar{p}}$ = $\Delta N_{p}$) 
distributions obtained by measuring the number of protons 
($N_{p}$)  and anti-protons ($N_{\bar{p}}$)  at mid-rapidity ($|y|$ $<$ 0.5) in 
the transverse momentum range 0.4 $<$ \pt (GeV/$c$)$<$ 2.0 for
Au+Au collisions at various \rootsnn.
To study the shape of the 
event-by-event net-proton distribution in detail, cumulants ($C_{n}$) of various
orders are calculated, where $C_{1}$ = $M$, $C_{2}$ = $\sigma^2$,
$C_{3}$ = $S\sigma^{3}$ and $C_{4}$ = $\kappa\sigma^{4}$.

\begin{figure}[t!]
{\centerline{\includegraphics[width=0.5\textwidth]{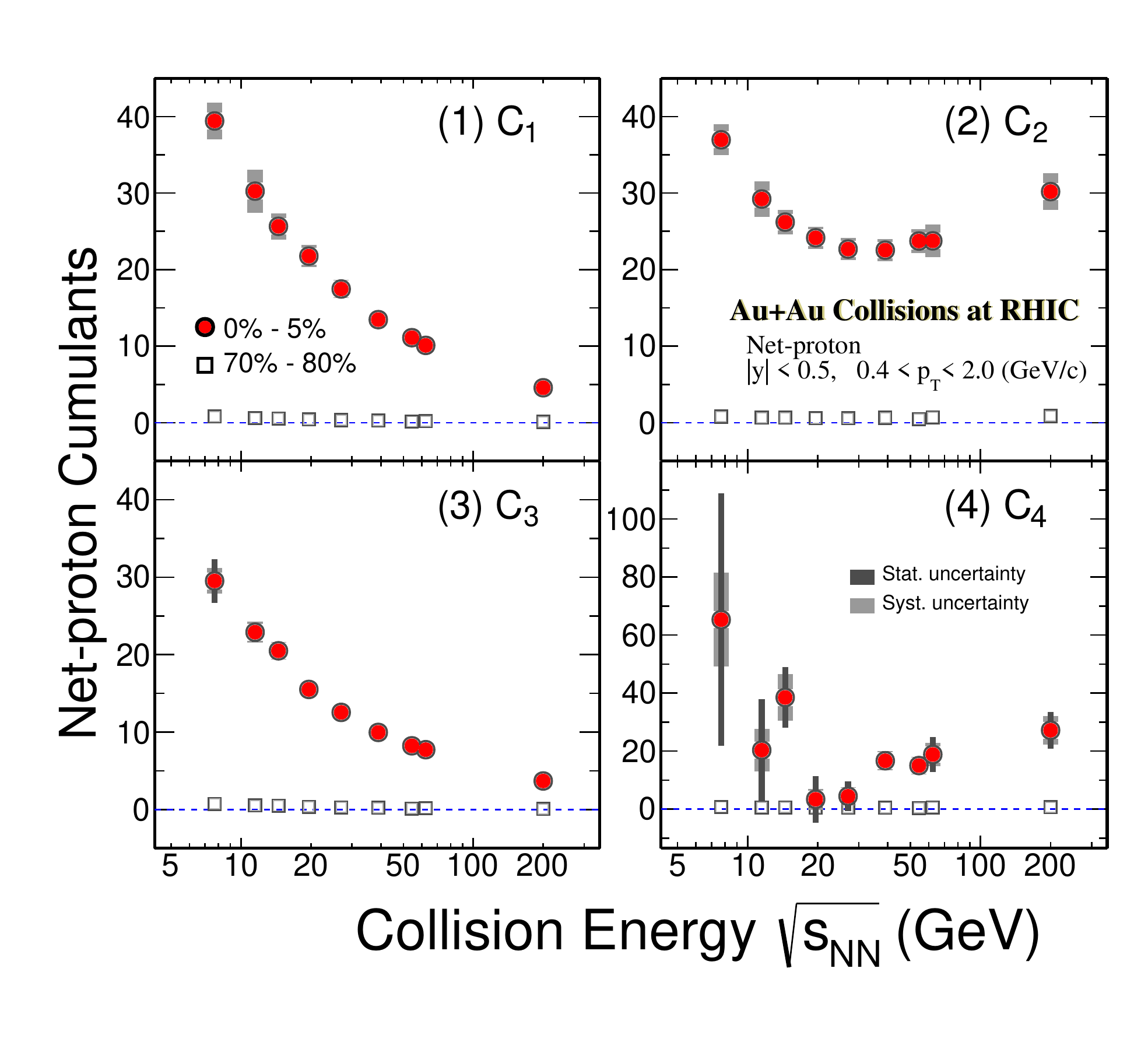}}}
\caption{
\label{fig2:Cumulants}
Cumulants ($C_{n}$) of the net-proton distributions for central (0-5\%)
and peripheral (70-80\%) Au+Au 
collisions  as a function of collision energy. The transverse
momentum (\pta) range for the
measurements is from 0.4 to 2 GeV/$c$ and the rapidity ($y$) range
is -0.5 $<$ $y$ $<$ 0.5.
}
\end{figure}

Figure~\ref{fig2:Cumulants} shows the net-proton
cumulants ($C_{n}$) as a function of \rootsnn for central and peripheral (see Supplemental Material~\cite{sup} for a magnified version). Au+Au
collisions. The cumulants are
corrected for the multiplicity variations arising due to finite
impact parameter range for the
measurements~\cite{Luo:2013bmi}. These corrections suppress the volume
fluctuations considerably~\cite{Luo:2013bmi,Sugiura:2019toh}.
A different volume fluctuation correction method~\cite{Braun-Munzinger:2016yjz} has been
applied to the 0-5\% central Au+Au collision data and the results were
found to be consistent with those shown in Fig~\ref{fig2:Cumulants} .
The cumulants are also
corrected for finite track reconstruction efficiencies of the TPC and
TOF detectors.
This is done by assuming a binomial response of the two detectors~\cite{Adamczyk:2013dal,Luo:2014rea}.
A cross-check using a different 
method based on unfolding~\cite{sup} of the distributions for central Au+Au collisions at \rootsnn = 200 GeV has been found to give values consistent with the cumulants shown in
Fig.~\ref{fig2:Cumulants}. Further, the efficiency
  correction method used has been
verified in a Monte Carlo calculation.
Typical values for the
efficiencies in the TPC (TOF-matching)
for the momentum range studied in 0-5\% central Au+Au collisions at \rootsnn =
7.7 GeV are 83\%(72\%) and 81\%(70\%) for the protons and
anti-protons, respectively. The corresponding efficiencies for \rootsnn =
200 GeV collisions are 62\%(69\%) and 60\%(68\%) for the protons and
anti-protons, respectively.  The statistical uncertainties are obtained using
both a bootstrap approach~\cite{Pandav:2018bdx,Luo:2014rea} and the Delta theorem~\cite{Pandav:2018bdx,Luo:2014rea, Luo:2011tp}
method. The systematic uncertainties are estimated by
varying the experimental requirements to reconstruct $p$ 
(${\bar{p}}$) in the TPC and TOF. These requirements include the distance of the proton
and anti-proton tracks from the primary vertex position,
track quality reflected by the number of TPC space
points used in the track reconstruction, the particle
identification criteria passing certain selection criteria, and the uncertainties in estimating the
reconstruction efficiencies. The systematic uncertainties at different
collision energies are uncorrelated.

The large values of $C_{3}$ and $C_{4}$ for central
Au+Au collisions show that the distributions have
non-Gaussian shapes,
a possible indication of enhanced fluctuations
arising from a possible critical point~\cite{Stephanov:2008qz,Asakawa:2009aj}.
The corresponding values for peripheral collisions are small and close to
zero. For central collisions, the $C_{1}$ and $C_{3}$
monotonically decrease with increasing \rootsnn.

We employ ratios of cumulants in order to cancel volume variations to first order.
Further, these ratios of cumulants are related to
the ratio of baryon-number
susceptibilities. The latter are $\chi^{\mathrm B}_{n}$ =  $\frac{d^nP}{d\mu_B^n}$,  
where $n$ is the order and $P$ is the pressure of the system at a given $T$ and $\mu_{\mathrm
  B}$, 
computed in lattice QCD and QCD-based models ~\cite{Gavai:2010zn}. The
$C_{3}/C_{2}$ = ${\it{S}}\sigma$ = $(\chi^{\mathrm B}_{3}/T)/(\chi^{\mathrm B}_{2}/T^2)$ and
$C_{4}/C_{2}$ = $\kappa\sigma^2$ = $(\chi^{\mathrm
  B}_{4})/(\chi^{\mathrm B}_{2}/T^2)$. Close to
the critical point, QCD-based calculations predict the net-baryon number distributions to be non-Gaussian and 
susceptibilities to diverge, causing moments, especially higher-order
quantities like $\kappa\sigma^{2}$, to have non-monotonic variation
as a function of \rootsnn~\cite{Gavai:2010zn,Stokic:2008jh}.

\begin{figure}[t!]
{\centerline{\includegraphics[scale=0.4]{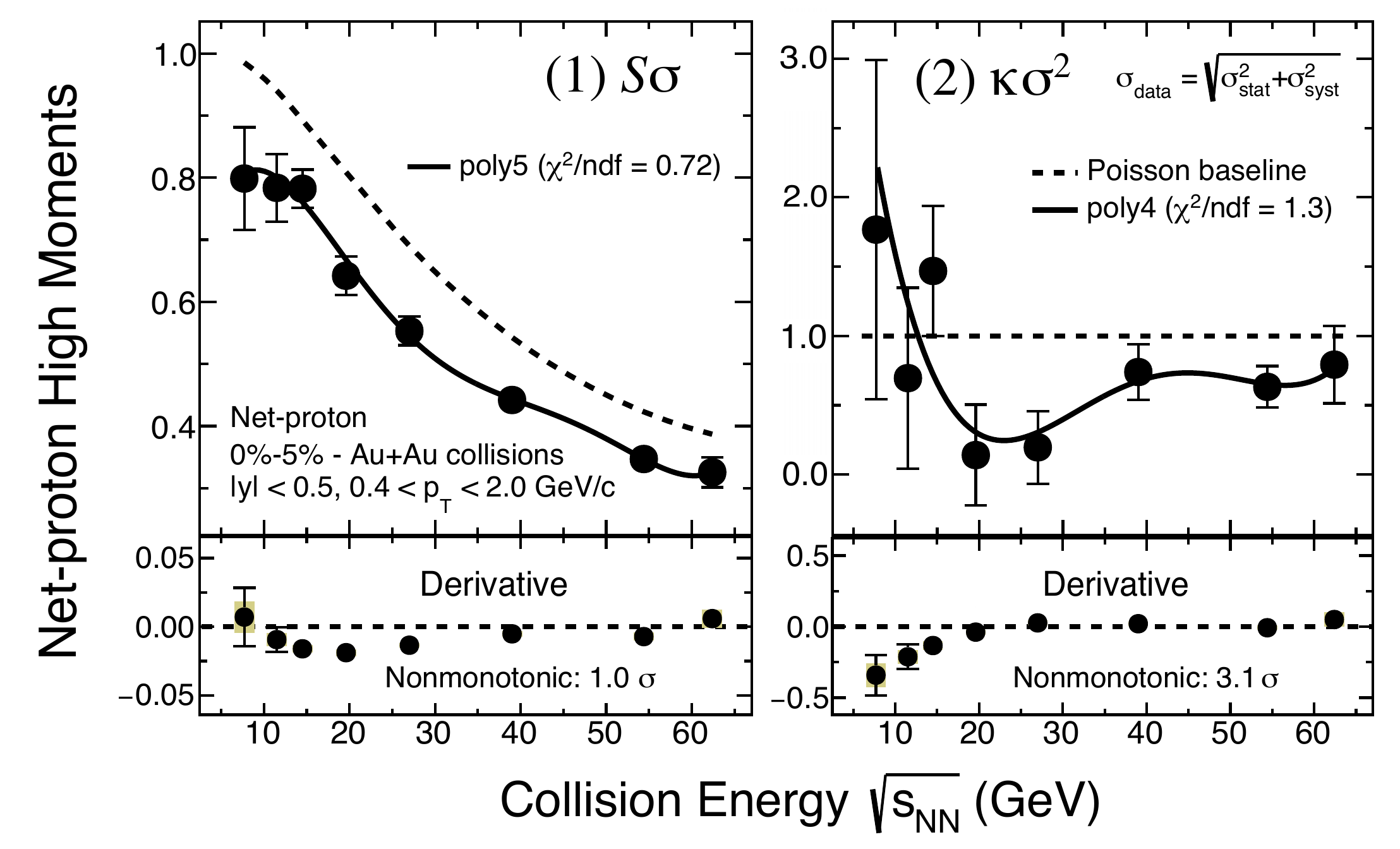}}}
\caption{ \label{fig3:nonmonotonic}
Upper panels: $\it{S}\sigma$ (1) and 
$\kappa\sigma^{2}$ (2) of net-proton distributions for 0-5\% central Au+Au 
                                collisions from \rootsnn = 7.7 -
                                62.4 GeV. The bar on the data points 
                                are statistical and systematic 
                                uncertainties added in quadrature. The 
                                black solid lines are polynomial fit 
                                functions which best describes the 
                                data. The black dashed 
                                lines are the Poisson baselines. 
                                Lower panels: Derivative of the fitted 
                                polynomial as a function of 
                                \rootsnn. The bar and the shaded band on 
                                the derivatives represent the statistical and systematic uncertainties, respectively.}
\end{figure}

Figure~\ref{fig3:nonmonotonic} shows the central 0-5\% Au+Au collision data for $\it{S}\sigma$
and $\kappa\sigma^{2}$ in the collision energy range of 7.7 -- 62.4
GeV, fitted to a polynomial function of order five and four,
respectively.
The derivative of the polynomial function changes sign~\cite{sup} with \rootsnn for
$\kappa\sigma^{2}$, thereby indicating a non-monotonic variation of the measurement with the 
collision energy. The uncertainties of the derivatives are obtained by varying the data 
points randomly at each energy within the statistical and systematic uncertainties separately. 
The overall significance of the change in the sign of the slope for $\kappa\sigma^{2}$
versus \rootsnn, based on the fourth order polynomial function fitting 
procedure from \rootsnn = 7.7 to 62.4 GeV, is 
3.1$\sigma$.
This significance is obtained by generating one million 
sets of points, where for each set, the measured $\kappa\sigma^{2}$ value at a given \rootsnn is randomly varied within the 
total Gaussian uncertainties (systematic and statistical uncertainties added in quadrature). Then 
for each new $\kappa\sigma^{2}$ versus \rootsnn set of points, a fourth order polynomial function is fitted and 
the derivative values are calculated at different \rootsnn (as discussed 
above). A total of 1143 sets were found to have 
the same derivative sign at all \rootsnn. The probability that at least one derivative 
at a given \rootsnn has a different sign
is found 
to be 0.998857, which corresponds to 3.1$\sigma$. 
A similar procedure was applied to the lower-order product of moments.
The $\sigma^{2}/M$ (not shown) strongly favors a monotonic energy dependence
excluding the non-monotonic trend at a 3.4 $\sigma$ level.
Within 1.0 $\sigma$ significance the $\it{S}\sigma$ allows for a
non-monotonic energy dependence.
This is consistent with a QCD based model expectation that the
higher the order of the moments the more sensitive it is to physics processes such as a 
critical point~\cite{Stephanov:2008qz}.

\begin{figure*}[t!]
{\centerline{\includegraphics[scale=0.65]{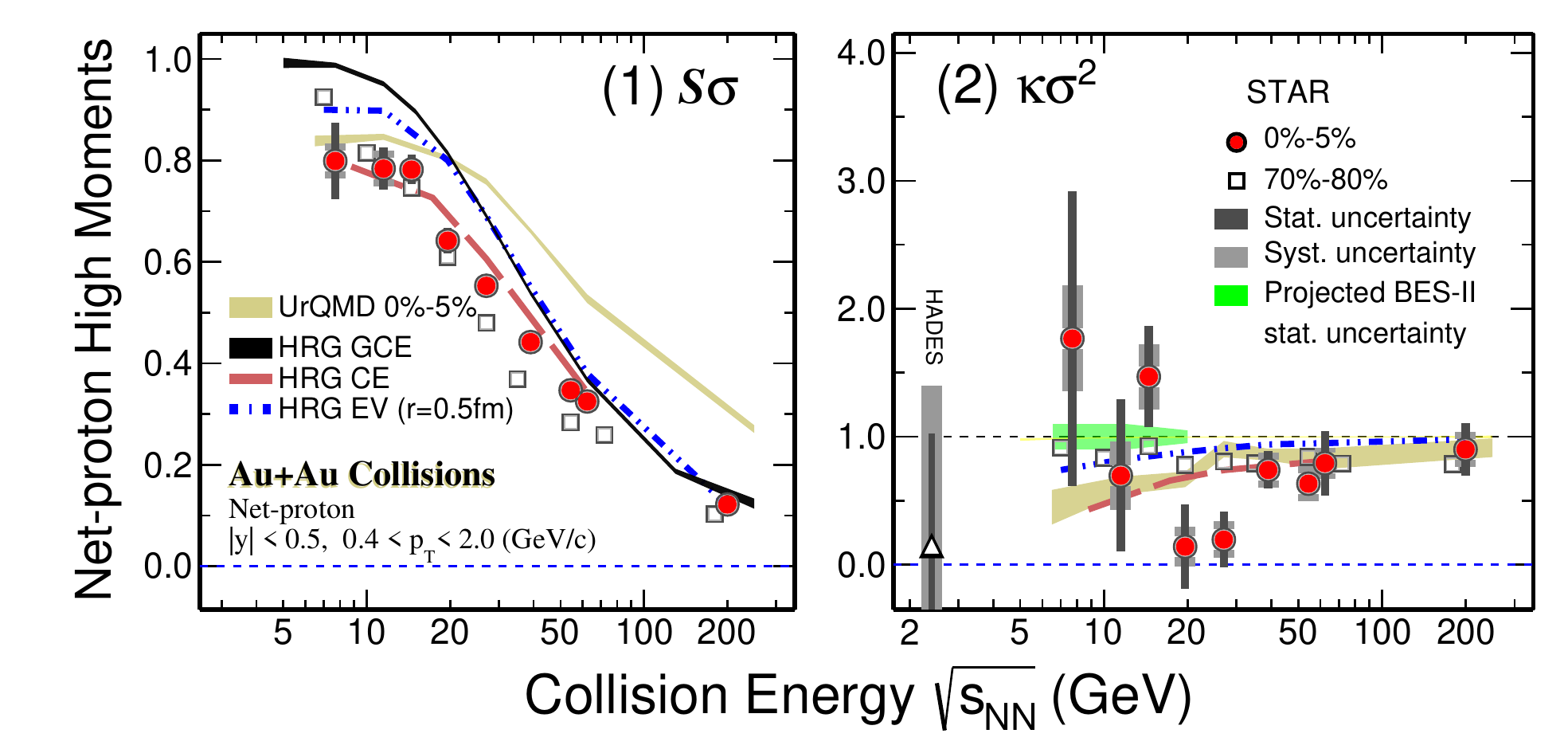}}}
\caption{ \label{fig4:kurtosis}
$\it{S}\sigma$ (1) and $\kappa\sigma^{2}$ (2) 
as a function of collision energy for net-proton distributions 
measured in Au+Au collisions. The results  are shown for central 
(0-5\%, filled circles ) and peripheral (70-80\%, open squares) collisions within 0.4 $<$ \pt 
(GeV/$c$) $<$ 2.0 and $|y|$ $<$ 0.5. 
The vertical narrow and wide bars represent the 
statistical and systematic uncertainties, respectively. Shown as an open
triangle is the result from the HADES experiment~\cite{Adamczewski-Musch:2020slf} for 0-10\% 
Au+Au collisions and $|y|$ $<$ 0.4. The shaded green band is the estimated statistical uncertainty for BES-II. 
The peripheral data points have been shifted along the $x$-axis for clarity of presentation. 
Results from different variants (GCE, EV, CE) of the hadron resonance gas (HRG) 
model~\cite{Garg:2013ata,Bhattacharyya:2013oya,Braun-Munzinger:2020jbk}
and a 
transport model calculation (UrQMD~\cite{Bleicher:1999xi}) for central 
collisions (0-5\%) are shown as black, red, blue bands and a gold band, respectively.
}
\end{figure*}

Figure~\ref{fig4:kurtosis} shows the variation of $\it{S}\sigma$
(or $C_{3}/C_{2}$) and  $\kappa\sigma^{2}$ (or
$C_{4}/C_{2}$) as a function of \rootsnn for central  
and peripheral Au+Au collisions. 
In central collisions, as discussed above, a non-monotonic variation with beam energy is
observed for $\kappa\sigma^2$.
The peripheral collisions on the other hand do not show a
non-monotonic variation with \rootsnn around the statistical baseline
of unity,  and $\kappa\sigma^2$ values are always below unity.
It is worth noting that in peripheral collisions, the 
system formed may not be hot and dense enough to undergo a phase
transition or come close to the QCD critical point.
The expectations from an ideal statistical model of hadrons assuming 
thermodynamical equilibrium, called the Hadron Resonance Gas (HRG) 
model~\cite{Garg:2013ata}, calculated within the experimental 
acceptance and considering a grand canonical ensemble (GCE), excluded
volume (EV)~\cite{Bhattacharyya:2013oya}, and canonical ensemble (CE)~\cite{Braun-Munzinger:2020jbk}, are also shown in 
Fig.~\ref{fig4:kurtosis}.
The HRG results do not quantitatively describe the data.
Corresponding $\kappa\sigma^{2}$ ($\it{S}\sigma$) results for 0-5\%
Au+Au collisions from a transport-based UrQMD model~\cite{Bleicher:1999xi} calculation, which 
incorporates conservation laws and most of the relevant physics apart 
from a phase transition or a critical point, and which is calculated 
within the experimental acceptance, show a monotonic decrease (increase) 
with decreasing collision energy (see Supplemental Material~\cite{sup} for a
quantitative comparison).
An exercise with the UrQMD and HRG model with canonical ensemble as
the non-critical baseline yielded a similar significance as 
reported in Fig.~\ref{fig3:nonmonotonic}.
Similar conclusions are obtained 
  from JAM~\cite{Zhang:2019lqz}, another microscopic transport model. 
Neither of the model calculations explains simultaneously the measured dependence of 
the $\kappa\sigma^{2}$ and $\it{S}\sigma$ of the net-proton distribution on \rootsnn for 
central Au+Au collisions. This can be seen from the values of a $\chi^2$ test between
the experimental data and various models for \rootsnn = 7.7 - 27 GeV
given in Table~\ref{tab1}, $p$ reflects the probability that a model
agrees with the data. However, for a wider energy range \rootsnn = 7.7
- 62.4 GeV the $p$ value with respect to HRG CE is larger than 0.05~\cite{Braun-Munzinger:2020jbk}.

\begin{table}
	\caption{The $p$ values of a $\chi^2$ test between data and
          various models for the \rootsnn dependence of $\it{S}\sigma$  and
          $\kappa\sigma^{2}$  values of net-proton
          distributions in 0-5\% central Au+Au collisions. The results
          are for the energy range 7.7 to 27 GeV which is relevant
          for the search for a critical point~\cite{Bazavov:2017tot, Bazavov:2017dus}.}
	\centering   
	\begin{tabular}{|c|c|c|c|c|}
		\hline	
		Moments & HRG GCE & HRG EV & HRG CE &
                                                                  UrQMD  \\
          	 &  &  (r = 0.5 fm)&&  \\
		\hline 
 $\it{S}\sigma$ & $<$ 0.001 &  $<$ 0.001 & 0.0754 & $<$ 0.001 \\
\hline 
$\kappa\sigma^{2}$ & 0.00553& 0.0145 & 0.0450 & 0.0221\\
\hline  
	\end{tabular}
	\label{tab1}
\end{table}

In conclusion, we have presented measurements of net-proton cumulant
ratios with the STAR detector at RHIC over a wide range of $\mu_{\mathrm B}$ (20 to
420 MeV) which are relevant to a QCD
critical point search in the QCD phase diagram.
We have observed a non-monotonic behavior
as a function of \rootsnn, in net-proton $\kappa\sigma^2$ in central
Au+Au collisions with a significance of 3.1$\sigma$ relative to
Skellam expectation. Other baselines without a critical point result in similar significance.
In contrast, monotonic behavior with \rootsnn is predicted for the
statistical hadron gas model, and for a nuclear transport model
without a critical point, as observed experimentally in peripheral collisions. 
The deviation of the measured $\kappa\sigma^2$ from several baseline
calculations with no critical point, and its non-monotonic dependence on
\rootsnn, are qualitatively  consistent with expectations from a QCD-based model which
includes a critical point~\cite{Stephanov:2008qz,Stephanov:2011pb}.
Our measurements can also be compared to 
the baryon-number susceptibilities computed from QCD to understand  
various other features of the QCD phase structure as well as to obtain 
the freeze-out conditions in heavy-ion collisions.
Higher event statistics will allow for a more differential measurement of 
experimental observables in $y$-$p_{\mathrm T}$. They will improve the comparison of the 
measurements with QCD calculations which include the dynamics associated 
with heavy-ion collisions, and hence they may help in establishing the 
critical point.

We thank P. Braun-Munzinger, S. Gupta, F. Karsch, M. Kitazawa, V. Koch,
D. Mishra, K. Rajagopal, K. Redlich, and M. Stephanov for stimulating discussions.
We thank the RHIC Operations Group and RCF at BNL, the NERSC Center at LBNL, and the Open Science Grid consortium for providing resources and support.  This work was supported in part by the Office of Nuclear Physics within the U.S. DOE Office of Science, the U.S. National Science Foundation, the Ministry of Education and Science of the Russian Federation, National Natural Science Foundation of China, Chinese Academy of Science, the Ministry of Science and Technology of China and the Chinese Ministry of Education, the Higher Education Sprout Project by Ministry of Education at NCKU, the National Research Foundation of Korea, Czech Science Foundation and Ministry of Education, Youth and Sports of the Czech Republic, Hungarian National Research, Development and Innovation Office, New National Excellency Programme of the Hungarian Ministry of Human Capacities, Department of Atomic Energy and Department of Science and Technology of the Government of India, the National Science Centre of Poland, the Ministry  of Science, Education and Sports of the Republic of Croatia, RosAtom of Russia and German Bundesministerium fur Bildung, Wissenschaft, Forschung and Technologie (BMBF), Helmholtz Association, Ministry of Education, Culture, Sports, Science, and Technology (MEXT) and Japan Society for the Promotion of Science (JSPS).



\section{Supplemental material}

\subsection{ Event selection and proton and anti-proton identification in
  STAR detector}

To reject pile-up and 
other background events, information from the fast detectors, a scintillator based 
vertex position detector (VPD)~\cite{Llope:2003ti} and the time-of-flight (TOF)
detector~\cite{Llope:2003ti,Llope:2012zz} and the time projection chamber (TPC)~\cite{Anderson:2003ur} are
used. To further ensure a good quality of data, run
by run study of several 
variables was carried out to remove bad events. The variables
used include the total number of uncorrected charged particles,
average transverse momentum in an event,
mean pseudorapidity and azimuthal angle in an event etc. In addition,
the distance of closest approach (DCA) of the charged particle track
from the primary vertex, especially the signed transverse average DCA 
and its stability, are studied to remove bad events. These classes
of bad events are primarily related to the unstable beam conditions during the
data taking and improper space-charge calibration of the TPC.
The number of events for the top 5\% central collisions ranges between 0.14 million for
\rootsnn = 7.7 GeV and 33 million at \rootsnn = 54.4 GeV.

 \begin{figure}[t!]
{\centerline{\includegraphics[scale=0.4]{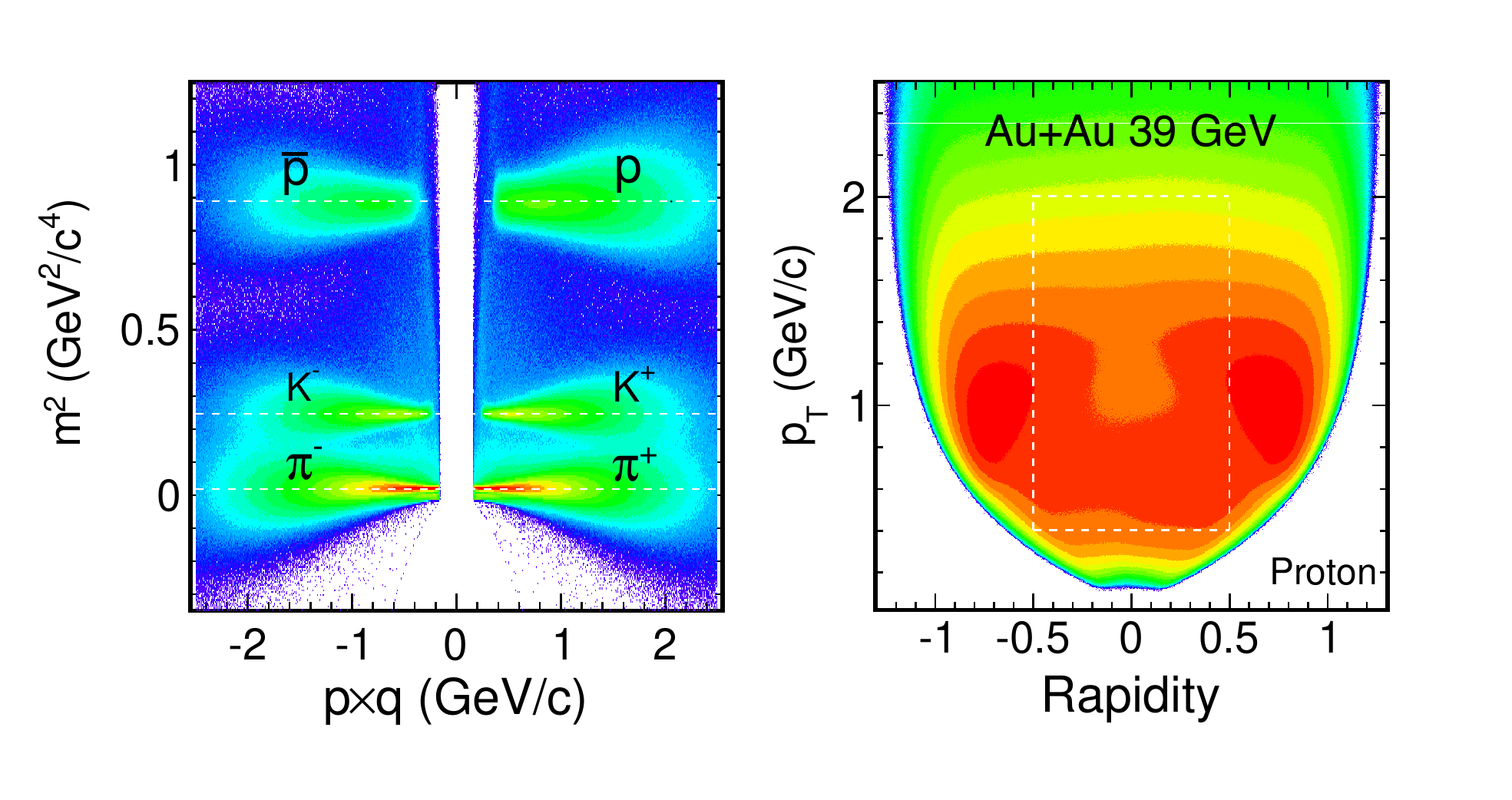}}}
\caption{
\label{fig:pid-accp}
Left panel: Square of the mass of the charged particles, requiring
timing information from the TOF, as a function 
of the product of the momentum ($p$) and the ratio of the particle’s
charge to the elementary charge $e$ ($q$),
both measured using the TPC  in Au+Au collisions at \rootsnn = 39
GeV. The white dashed lines correspond to the expected square of the mass 
of each particle species.
Right panel: The transverse momentum  (\pt) 
versus the rapidity ($y$) for protons measured in the STAR 
detector for Au+Au collisions.
}
\end{figure}

Figure~\ref{fig:pid-accp} (left panel) shows a typical distribution of
the square of the
mass associated with each track in an event obtained from the
TOF~\cite{Llope:2003ti,Llope:2012zz} as a function of the product of the momentum and
the charge of the track determined by the TPC~\cite{Anderson:2003ur}.
The proton candidates are well separated from other hadrons like kaons and pions. The right panel of 
Fig.~\ref{fig:pid-accp} shows \pt versus $y$
for  protons in the STAR detector. The white dashed rectangular box is
the region selected for the results presented here. It may be noted
that STAR, being a collider experiment, has a \pt versus $y$ 
acceptance near mid-rapidity that is uniform across all beam energies studied. Uniform 
acceptance allows for the results to be directly compared across all 
the \rootsnn.

The constant \pt versus $y$ acceptance near mid-rapidity raises the
issue of contribution of background protons to the analysis. This can
be gauged by looking at the DCA of the
proton tracks from the primary vertex and comparing it to the
corresponding results for the anti-protons.  A DCA criterion of less than 1
cm is used in the analysis reported here. This criterion
reduces the background proton contributions in the momentum range of
the study to less than 2-3\%. This small effect across all beam
energies is added to the systematic uncertainties obtained by varying
the DCA criteria between 1.2 and 0.8 cm.

\subsection{Efficiency corrections using unfolding of net-proton multiplicity 
distributions}

The unfolding method~\cite{Esumi:2020xdo} was applied to a data set that provides the most dense charged
particle environment in the detectors (0-5\% central Au+Au
collisions at \rootsnn = 200 GeV), where one expects the
maximum non-binomial detector effects. Detector-response matrices were
determined based on detector simulations with respect to generated and
measured protons and anti-protons~\cite{Nonaka:2019fhk}. All
possible non-binomial effects, including multiplicity dependent
efficiency, were corrected by utilizing the response matrices. The
detector response in such cases was found to be best described by a
beta-binomial distribution. Even in this situation, the
differences in the binomial~\cite{Nonaka:2017kko}  and unfolding methods of efficiency
correction were at a level of less than one $\sigma$ of the uncertainties.

\begin{figure*}[htp]
\includegraphics[scale=0.7]{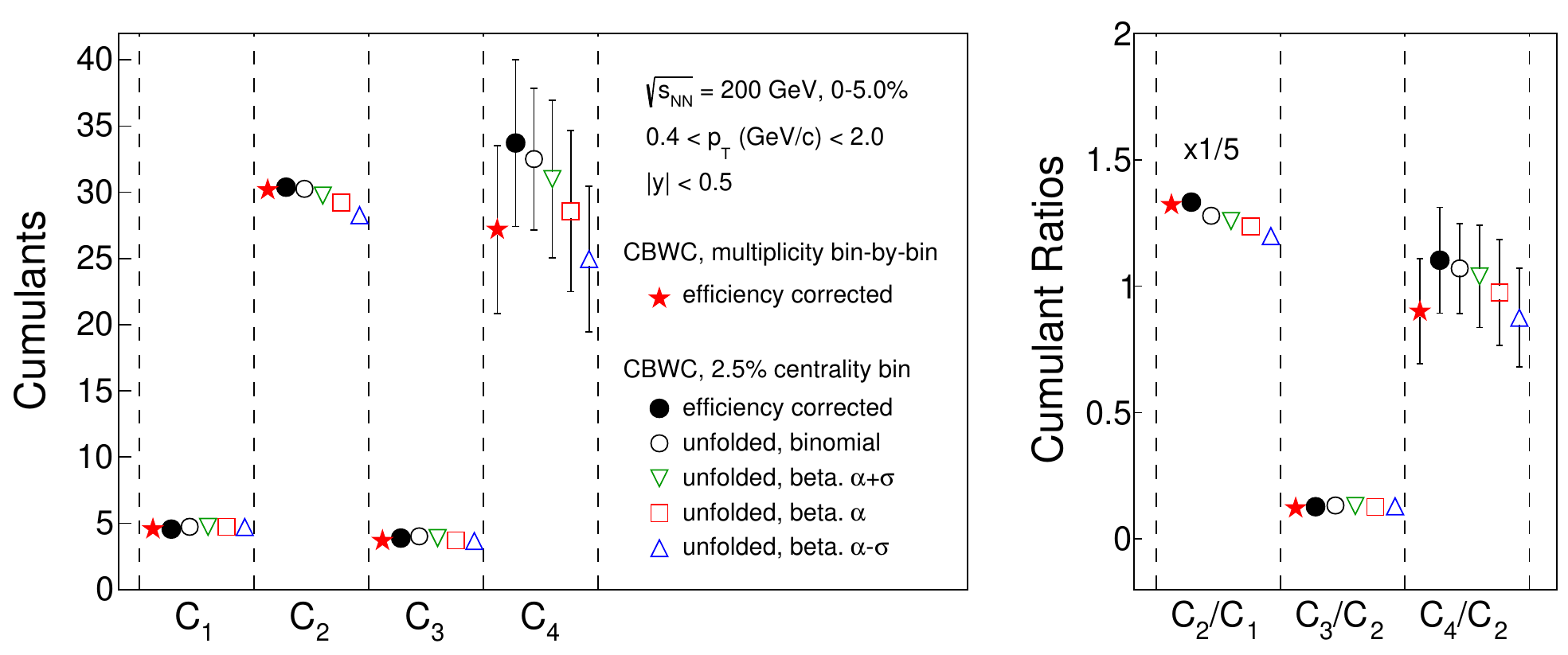}
\caption 
{Cumulants and their ratios up to the fourth order, 
  corrected for proton and anti-proton reconstruction efficiencies 
		in \rootsnn =~200~GeV Au+Au collisions at 0-5\% centrality. 
		Results from the conventional efficiency correction are shown as black filled circles, 
		results from the unfolding with the binomial detector 
                response are shown as black open circles, 
		and results from beta-binomial detector response with $\alpha+\sigma$, $\alpha$ 
		and $\alpha-\sigma$ are shown as green triangles, red 
                squares and blue triangles, respectively. The 
                parameter $\alpha$ quantifies the deviation from 
                binomial effects, obtained from simulation. 
		$C_{2}/C_{1}$ is scaled by a constant
                factor. For the data presented as red stars, the centrality
                  bin width correction (CBWC) is
                  applied for each multiplicity bin of the
                  multiplicity distribution used for centrality
                  determination (the result presented in the paper), while
                  for the other results, the cumulants are calculated
                  in centrality bins of width 2.5\% and averaged to
                  0-5\% centrality. This procedure was 
                  because the unfolding approach requires large
                  statistics, and thus is difficult toapply to each
                  multiplicity bin.
}
\label{fig:UnfFinal}
\end{figure*}

Cumulants and their ratios up to the fourth order, corrected for the
detector efficiencies using the unfolding method, are shown in
Fig.~\ref{fig:UnfFinal} for 0-5\% central Au+Au collisions at \rootsnn
= 200 GeV. The results are obtained by using centrality bin width correction
(CBWC)~\cite{Luo:2013bmi} at 2.5\% bin width. For each column, the
first point is efficiency corrected using the binomial model method
(as employed in the present analysis), the next point is the result corrected for  
the binomial detector response using the unfolding technique, and the last three points are from 
unfolding using the beta-binomial response with three values of the
non-binomial parameter. 
The results are ordered from left to right in terms of increasing
deviations of the response function compared to the binomial distribution.
Checks using unfolding of the distributions for central Au+Au collisions have been found to yield values consistent with the
cumulants obtained using the default binomial method of efficiency correction,
within the current statistics of the measurements. An alternate
approach called the moment expansion method~\cite{Nonaka:2018mgw} was used for efficiency
correction and found to be consistent with the unfolding method.

\subsection{Cumulants for 70-80\% Au+Au collisions net-proton
  distribution}

\begin{figure}[htp]
\centerline{\includegraphics[scale=0.4]{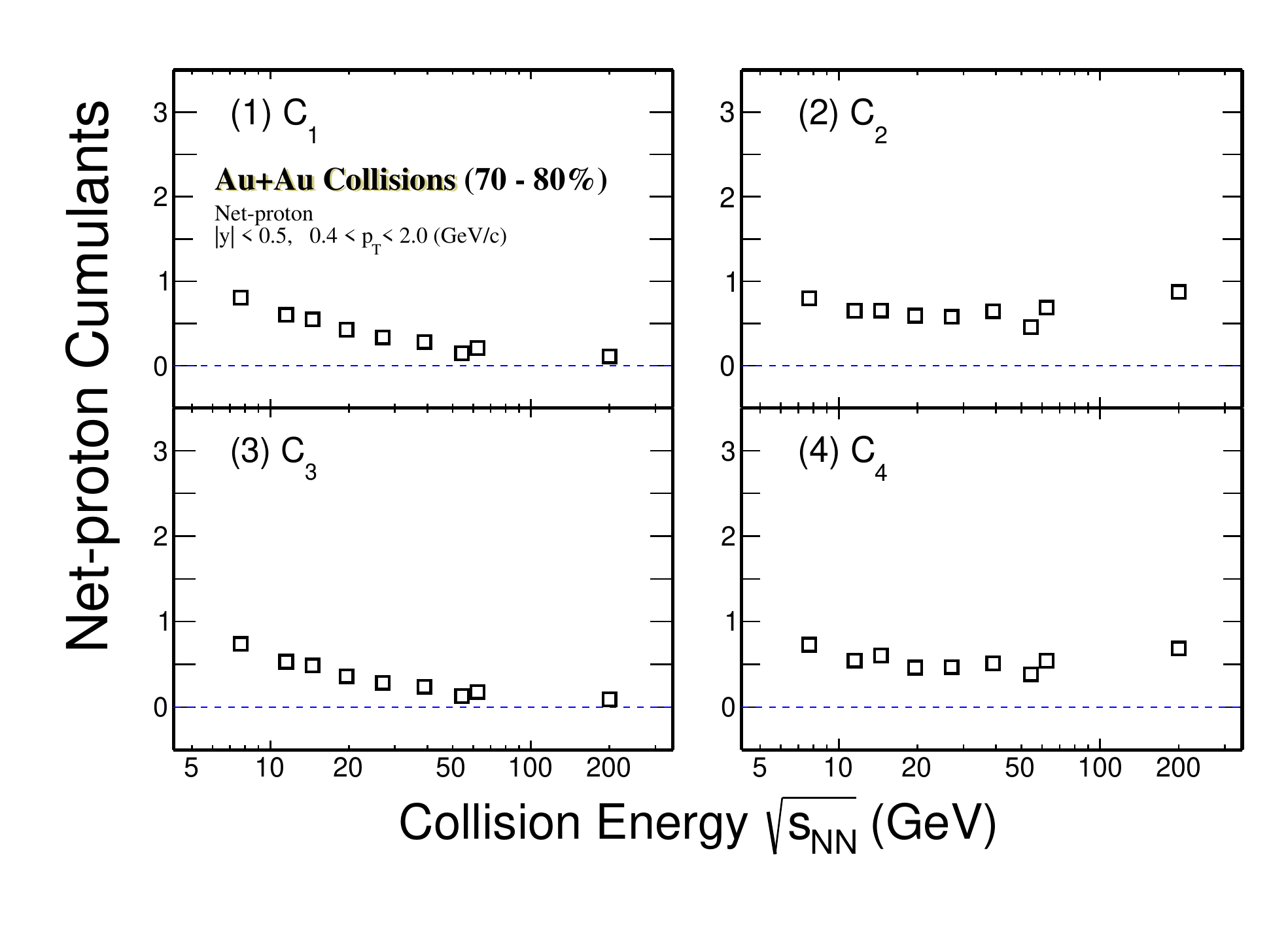}}
\caption{
Cumulants ($C_{n}$) of the net-proton distributions for peripheral (70-80\%) Au+Au 
collisions  as a function of collision energy. The transverse
momentum (\pta) range for the
measurements is from 0.4 to 2 GeV/$c$ and the rapidity ($y$) range
is -0.5 $<$ $y$ $<$ 0.5. The uncertainties are small and within the
symbol size. }
\label{fig:peri}
\end{figure}
Figure~\ref{fig:peri} shows a magnified version of the peripehral
(70-80\%) Au+Au collisions data presented in Fig.2 of the paper.

\subsection{Rapidity dependence of $C_{4}$/$C_{2}$ for 0-5\% central
  Au+Au collisions}
\begin{figure*}[htp]
\centerline{\includegraphics[scale=0.6]{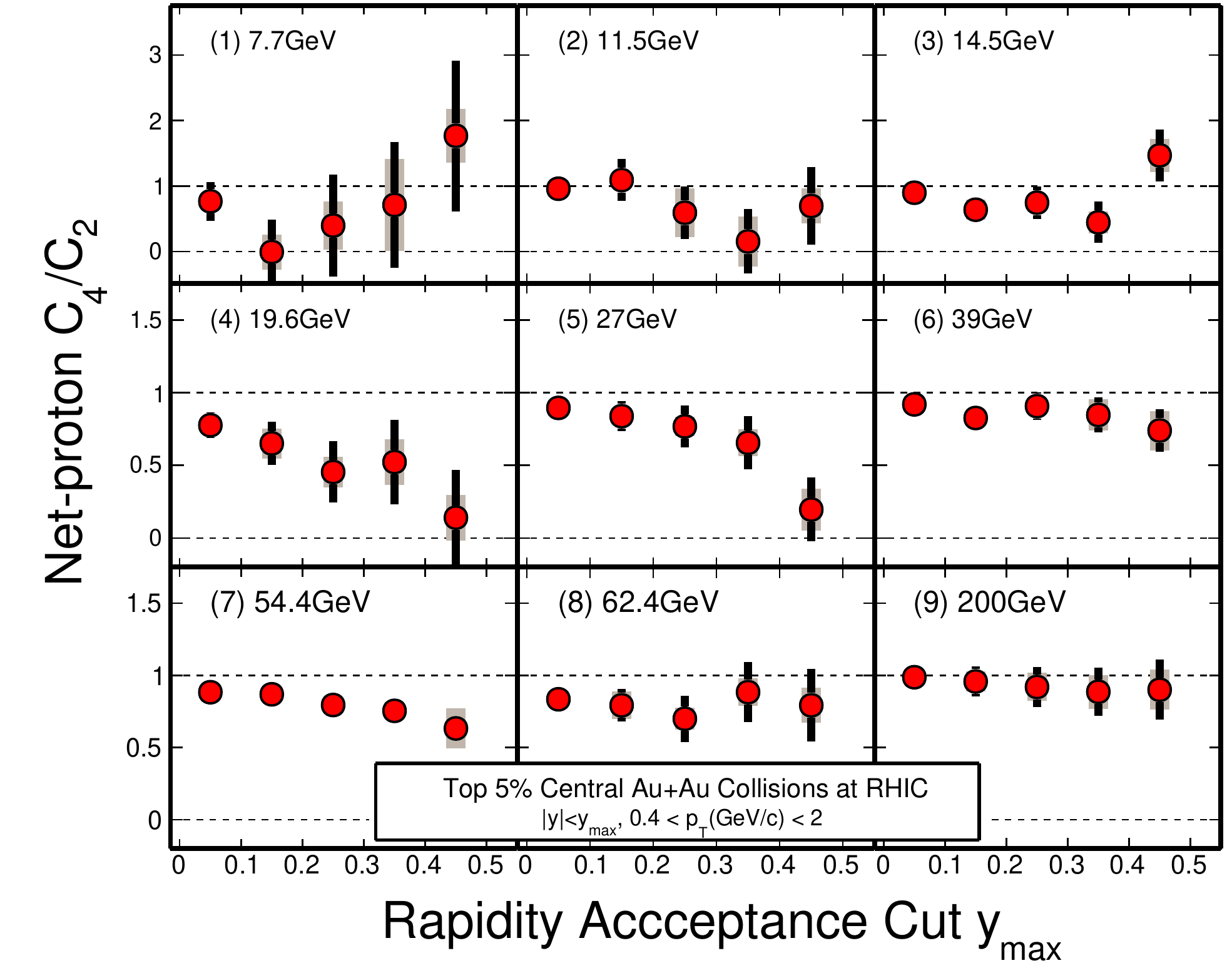}}
\caption{
Rapidity dependence of cumulant ratio $C_{4}/C_{2}$ of net-proton multiplicity distributions 
for 0-5\% central Au+Au collisions at \rootsnn = 7.7, 11.5, 14.5, 
19.6, 27, 39, 54.4, 62.4, and 200 GeV. The 
measurements are done for the $p_{T}$ range of 0.4 to 2.0 GeV/$c$. The 
lines and shaded areas represent 
statistical and systematic uncertainties. }
\label{fig:rap}
\end{figure*}

The cumulant ratio $C_{4}/C_{2}$ of net-proton multiplicity distributions
for 0-5\% central Au+Au collisions at \rootsnn = 7.7, 11.5, 14.5,
19.6, 27, 39, 54.4, 62.4, and 200 GeV is shown in Fig.~\ref{fig:rap}.
The $C_{4}/C_{2}$ value is close to unity for all collision energies for
the smallest rapidity acceptance. At \rootsnn = 200 GeV, the
$C_{4}/C_{2}$ values remain close to unity as rapidity
acceptance is increased, while for \rootsnn = 7.7 GeV, the
$C_{4}/C_{2}$ values first shows a drop followed by a
marginally significant increase as rapidity acceptance is increased. The $C_{4}/C_{2}$
values decrease as rapidity acceptance is increased at the intermediate
collision energies of \rootsnn = 19.6 and 27 GeV.

\subsection{Deviation in $\kappa\sigma^{2}$ values at various \rootsnn
  of 0-5\% central collision data from models and 70-80\% peripheral collisions:}
\begin{figure}[htp]
\centerline{\includegraphics[scale=0.5]{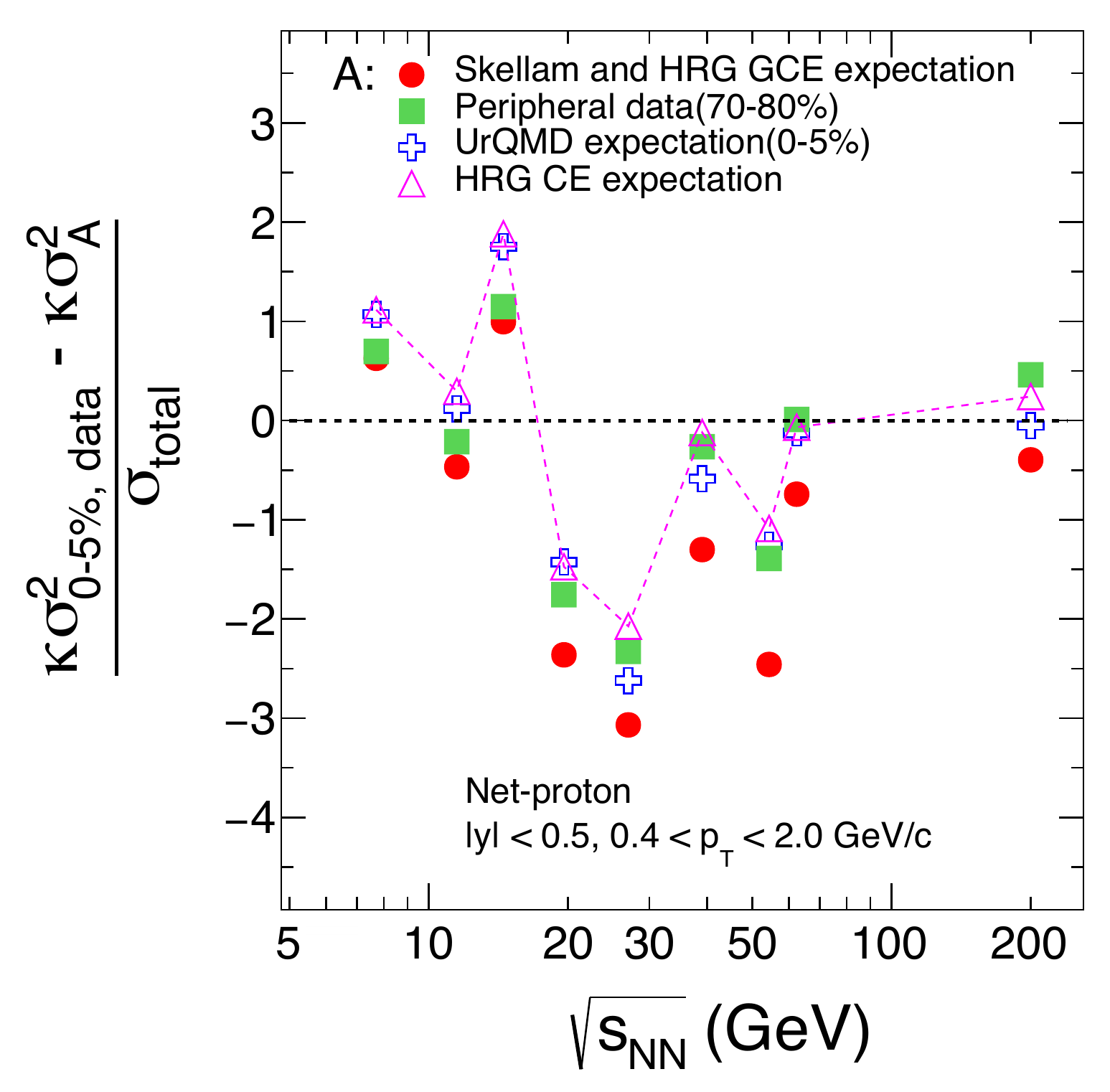}}
\caption{Significance of deviation in net-proton distribution $\kappa\sigma^{2}$ values for central 0-5\% Au+Au
collisions and those from UrQMD, HRG (grand canonical ensemble), HRG
(canonical ensemble) and 70-80\% peripheral Au+Au
collisions. The difference between data and model is divided by the
total uncertainties to obtain the significance. The
results are plotted as a function of \rootsnn. The $\sigma_{\rm
  {total}}$ =  $\sqrt{ {\sigma^{2}_{\rm {total, 0-5\%}}}  + {\sigma^{2}_{\rm
  {total, A}}}}$. The  $\sigma_{\rm  {total, 0-5\%}}$ is the statistical and
systematic uncertainties on  $\kappa\sigma^{2}$ values for central 0-5\% Au+Au
collisions added in quarature. The $\sigma_{\rm {total, A}} $
  is the statistical uncertantity from  models (UrQMD or HRG) or the statistical and
systematic uncertainties on  $\kappa\sigma^{2}$ values for 70-80\%
peripheral Au+Au collisions added in quadrature. The HRG (canonical
ensemble) results are connected by a dashed line.}
\label{fig:dev}
\end{figure}

{Figure~\ref{fig:dev} shows the deviation of $\kappa\sigma^{2}$ values for central 0-5\% Au+Au
collisions from the corresponding values from UrQMD and HRG
models. Also shown is the deviation from 70-80\% peripheral Au+Au
collisions. The uncertainties used to obtain the deviations are
statistical and systematic added in quadrature. The central collision
data deviates qualitatively in a similar manner for all the baseline
measures as a function of \rootsnn. The deviations are both positive
and negative in sign.

\subsection{Values of polynomial function fit to  $\kappa\sigma^{2}$
  and $\it{S}\sigma$ and their derivatives versus \rootsnn:}

\begin{table}
  \caption{Values of the parameters of fourth (fifth) order polynomial 
  that describes the collision energy dependence of $\kappa\sigma^{2}$
  ($\it{S}\sigma$) at various \rootsnn along with their 
uncertainties. The polynomials are of the form $\sum_{n} p_{n}
({\sqrt{s_{\rm NN}}})^{n}$, n = 0-4 for fourth order polynomial and 0-5 
for fifth order polynomial and $p_{n}$ are the parameters.}
	\centering   
	\begin{tabular}{|c|c|c|}
			\hline	
		Parameters &  $\kappa\sigma^{2}$&$\it{S}\sigma$    \\
		\hline 
		$p_{0}$& 6.24 $\pm$ 1.78 &0.51 $\pm$ 0.46 \\
		\hline 
		$p_{1}$ & --0.72 $\pm$ 0.22 &0.08 $\pm$ 0.09 \\
		\hline 
		$p_{2}$ & 0.03$\pm$0.01  &--0.007$\pm$ 0.006 \\
		\hline 
		$p_{3}$ & --0.0005$\pm$0.0002 & 0.0002 $\pm$ 0.0002\\
		\hline 
		$p_{4}$ & 0.000003 $\pm$ 0.000001& --3.3$\times10^{-6}$ $\pm$ 2.7$\times10^{-6} $\\
		\hline 
		$p_{5}$ & -- & 1.8$\times10^{-8}$ $\pm$ 1.5$\times10^{-8} $\\
		\hline 
	\end{tabular}
	\label{tab1}
\end{table}

The values of the parameters of the polynomial functions for $\kappa\sigma^{2}$
and $\it{S}\sigma$ at \rootsnn = 7.7, 11.5, 14.5, 19.6, 27, 39, 54.4 and
62.4 GeV are given in Table~\ref{tab1}. The uncertainties on the
parameters are from the fitting procedure taking into account both the
statistical and systematic uncertainties on the data. The
$\chi^{2}$/NDF = 1.3 for the fourth order polynomial fit to
$\kappa\sigma^{2}$  versus \rootsnn and the $\chi^{2}$/NDF = 0.72 for
the fifth order polynomial fit to $\it{S}\sigma$ versus \rootsnn.
The \rootsnn =
200 GeV data point is not included to quantify the non-monotonic
variations as the polynomial function fits either did not converge or
yielded larger $\chi^{2}$ values. It may also be noted
that the possible critical point is predicted to exist at baryon
chemical potential values much larger than those corresponding to \rootsnn
= 200 GeV.

\begin{table*}
  \caption{Values of the derivative of the fourth (fifth) order polynomial that describes the collision energy dependence of $\kappa\sigma^{2}$
($\it{S}\sigma$) at various \rootsnn. The first uncertainty on the 
derivative corresponds to statistical uncertainty on the data
points, the second uncertainty corresponds to systematic uncertainty on the data points assuming
they are fully correlated and the third uncertainty corresponds to the systematic uncertainty on the
data points assuming they are fully uncorrelated. Also shown is the significance of the difference from zero of each derivative.}
	\centering   
	\begin{tabular}{|c|c|c|c|c|}
		\hline	
		\rootsnn (GeV) & Derivative of polynomial ($\kappa\sigma^{2}$)&Sig.  & Derivative of polynomial ($\it{S}\sigma$)&Sig.  \\
		\hline 
		7.7 & --0.341$\pm$ 0.142 $\pm$ 0.031 $\pm$ 0.079  &2.1 &0.0071 $\pm$ 0.0214 $\pm$0.0054 $\pm$ 0.0111& 0.3\\
		\hline 
		11.5 & --0.212 $\pm$ 0.087 $\pm$ 0.022 $\pm$ 0.045 &2.2& --0.0094 $\pm$ 0.0088 $\pm$0.0029 $\pm$ 0.0044&1.0 \\
		\hline 
		14.5 & --0.133 $\pm$ 0.055 $\pm$ 0.016 $\pm$ 0.026 &2.2&--0.0161$\pm$0.004$\pm$  0.0014 $\pm$ 0.0024&3.5 \\
		\hline 
		19.6 & --0.039 $\pm$ 0.023 $\pm$ 0.009 $\pm$ 0.013 & 1.5&--0.0189 $\pm$0.0031  $\pm$0.0001 $\pm$0.002& 5.1\\
		\hline 
		27 & 0.026 $\pm$ 0.019 $\pm$ 0.002 $\pm$ 0.014 & 1.1&--0.0135$\pm$ 0.0017$\pm$ 0.0004 $\pm$ 0.0013& 6.4\\
		\hline 
		39 & 0.02$\pm$ 0.011 $\pm$ 0.001 $\pm$ 0.01 & 1.4&--0.0052$\pm$0.0022$\pm$0.0005 $\pm$ 0.0017& 1.9\\
		\hline 
		54.4 & --0.008 $\pm$ 0.018 $\pm$ 0.001 $\pm$ 0.011 & 0.4&--0.0072$\pm$0.0026$\pm$0.0001$\pm$ 0.0024&2.0 \\
		\hline 
		62.4 & 0.05 $\pm$0.058 $\pm$ 0.002  $\pm$ 0.047 &0.7 &0.0059 $\pm$0.007$\pm$ 0.0025$\pm$ 0.0062& 0.6\\
		\hline 
	\end{tabular}
	\label{tab2}
      \end{table*}
      
The values of the derivatives of the polynomial functions for $\kappa\sigma^{2}$
and $\it{S}\sigma$ for 0-5\% central Au+Au collisions at \rootsnn = 7.7, 11.5, 14.5, 19.6, 27, 39, 54.4 and
62.4 GeV are given in Table~\ref{tab2}. The uncertainties on the
derivatives are obtained by varying the data
points randomly at each energy
within the statistical and systematic uncertainties separately. This
process assumes that the systematic uncertainties on the data points are fully
uncorrelated. In addition, we also provide an estimate of systematic
uncertainty on the derivative at each \rootsnn which assumes the
systematic uncertainties on the data points to be fully correlated. The statistical uncertainties on the
derivative values are obtained by the random sampling of the
data points using a Gaussian distribution whose mean is the
$\kappa\sigma^{2}$ or $\it{S}\sigma$ value of the data and the width of the
Gaussian is the statistical uncertainty, for
the data point at each collision energy. The uncorrelated systematic
uncertainties are obtained in the same way. This results in a new
collision energy dependence of $\kappa\sigma^{2}$ and
$\it{S}\sigma$. This new set of data is then fitted to the same order
polynomial function as the default case and the derivative is obtained at each collision
energy. This process is repeated until the width of the distribution of
derivative values at each collision energy converges. The width of this distribution is taken as
the uncertainty on the derivative value. 
For obtaining the fully correlated systematic uncertainty on the
derivative value, all the $\kappa\sigma^{2}$  or $\it{S}\sigma$ data
points are shifted up or down by the systematic uncertainties
together.  Then the resultant collision energy dependence of
$\kappa\sigma^{2}$ or $\it{S}\sigma$ is fitted by the same order polynomial
function as the default case and derivative values obtained. The difference in the
derivative values from the default values is taken as the correlated
systematic uncertainty on the derivative values. Also shown in the Table~\ref{tab2} are the significance
values for the derivative to be non-zero at each \rootsnn, calculated using the
statistical and the uncorrelated systematic uncertainties added in
quadrature.

A typical critical point signal expected from theoretical prediction is
an oscillating pattern around the statistical baseline
($\kappa\sigma^{2}$ = 1)~\cite{Stephanov:2011pb}.  As the
$\kappa\sigma^{2}$ values for the peripheral 70-80\% Au+Au collisions
are always below the statistical baseline of unity, a test for a non-monotonic
variation study is not carried out. Further, the polynomial fits to
the peripheral data yield much larger $\chi^{2}$/NDF,  and for a
polynomial of order four, the fit does not converge.

\begin{table}
	\caption{Values of the derivative of fourth order polynomial
		that describes the $\kappa\sigma^{2}$ versus $M/\sigma^{2}$ ($C_{1}/C_{2}$ ). The first uncertainty on the
		derivative corresponds to statistical uncertainty
		on the data points, the second uncertainty corresponds to systematic
		uncertainty on the data points assuming they are fully correlated and
		the third uncertainty corresponds to the systematic uncertainty on the
		data points assuming they are fully uncorrelated. Also shown is the significance of the difference from zero of each derivative.}
	\centering   
	\begin{tabular}{|c|c|c|c|c|}
		\hline	
		\rootsnn (GeV) & $C_{1}/C_{2}$& Derivative of polynomial ($\kappa\sigma^{2}$)&Sig.  \\
		\hline 
7.7 & 1.067&14.967 $\pm$ 13.12 $\pm$ 1.749 $\pm$ 6.965  &1.0 \\
\hline 
11.5 & 1.035$  $&12.17 $\pm$ 9.109 $\pm$ 1.47 $\pm$ 4.625 &1.2 \\
\hline 
14.5 & 0.979&7.833 $\pm$ 4.114 $\pm$ 1.034 $\pm$ 1.869 &1.7 \\
\hline 
19.6 & 0.901&3.176 $\pm$ 1.953 $\pm$ 0.56 $\pm$ 1.315 & 1.3\\
\hline 
27 & 0.77&--1.365 $\pm$ 1.64 $\pm$ 0.072 $\pm$ 0.997 & 0.7\\
\hline 
39 & 0.597&--1.878$\pm$ 1.605 $\pm$ 0.055 $\pm$ 1.369 & 0.9\\
\hline 
54.4 &0.468& 1.116 $\pm$ 3.216 $\pm$ 0.164 $\pm$ 2.029 & 0.3 \\
\hline 
62.4 &0.425& 2.634 $\pm$6.618 $\pm$ 0.285  $\pm$ 4.225 &0.3 \\
\hline 
	\end{tabular}
	\label{tab3}
\end{table}
Various ansatz related to the fitting procedure have been checked to determine the robustness of the sign change of the derivative
values. These includes fitting the data to various orders of
polynomial function and varying the fitting range.
For example, the $\kappa\sigma^{2}$  versus \rootsnn is 
fitted to a third order polynomial, which yielded a
$\chi^{2}$/NDF = 1.6. The derivative values are found to be consistent
with those obtained by fitting the data using the polynomial of order
four. The significance of the non-monotonic variation of
$\kappa\sigma^{2}$  versus \rootsnn when fitted to third-order
polynomial is 2.1 $\sigma$. A systematic study of progressively
excluding lower and higher collision energy data points gives a
consistent derivative value as reported in the paper.
Further, as
suggested in Ref.~\cite{Bazavov:2017tot}, the $\kappa\sigma^{2}$ was plotted as a function of $M/\sigma^{2}$ to study
the sign change of the derivative values. The values of the
derivatives of the fourth order polynomial functions for $\kappa\sigma^{2}$
versus $M/\sigma^{2}$ are given in Table~\ref{tab3}. The conclusions of such
studies qualitatively remain the same as presented in the current
work. The significance of the non-monotonic variation of
$\kappa\sigma^{2}$ with $M/\sigma^{2}$ was found to be 3.1$\sigma$.

\begin{table*}
	\caption{Same as Table~\ref{tab2}, except that the probability 
        distribution for estimation of uncorrelated systematic 
        uncertainties is assumed to be a uniform distribution.}
	\centering   
	\begin{tabular}{|c|c|c|c|c|}
		\hline	
	\rootsnn (GeV) & Derivative of polynomial ($\kappa\sigma^{2}$)&Sig.  & Derivative of polynomial ($\it{S}\sigma$)&Sig.  \\
\hline 
7.7 & --0.341$\pm$ 0.142 $\pm$ 0.031 $\pm$ 0.045  &2.3 &0.0071 $\pm$ 0.0214 $\pm$0.0054 $\pm$ 0.0064& 0.3\\
\hline 
11.5 & --0.212 $\pm$ 0.087 $\pm$ 0.022 $\pm$ 0.026 &2.3& --0.0094 $\pm$ 0.0088 $\pm$0.0029 $\pm$ 0.0026&1.0 \\
\hline 
14.5 & --0.133 $\pm$ 0.055 $\pm$ 0.016 $\pm$ 0.015 &2.3&--0.0161$\pm$0.004$\pm$  0.0014 $\pm$ 0.0014&3.8 \\
\hline 
19.6 & --0.039 $\pm$ 0.023 $\pm$ 0.009 $\pm$ 0.008 & 1.6&--0.0189 $\pm$0.0031  $\pm$0.0001 $\pm$0.0012& 5.7\\
\hline 
27 & 0.026 $\pm$ 0.019 $\pm$ 0.002 $\pm$ 0.008 & 1.3&--0.0135$\pm$ 0.0017$\pm$ 0.0004 $\pm$ 0.0007& 7.3\\
\hline 
39 & 0.02$\pm$ 0.011 $\pm$ 0.001 $\pm$ 0.006 & 1.7&--0.0052$\pm$0.0022$\pm$0.0005 $\pm$ 0.001& 2.1\\
\hline 
54.4 & --0.008 $\pm$ 0.018 $\pm$ 0.001 $\pm$ 0.007 & 0.4&--0.0072$\pm$0.0026$\pm$0.0001$\pm$ 0.0014&2.4 \\
\hline 
62.4 & 0.05 $\pm$0.058 $\pm$ 0.002  $\pm$ 0.027 &0.8 &0.0059 $\pm$0.007$\pm$ 0.0025$\pm$ 0.0036& 0.8\\
\hline 
	\end{tabular}
	\label{tab4}
\end{table*}

As a cross check, we have estimated the uncorrelated systematic
uncertainties on the derivative values by assuming the probability
distribution for the uncertainties to be a uniform distribution. The
derivative value and the significance at each collision energy are shown in Table~\ref{tab4}. The significance
of the derivative values are increased compared to those shown in Table~\ref{tab2}.


\end{document}